\documentclass[aps, prapplied, twocolumn, superscriptaddress, tightenlines, footinbib]{revtex4-2}
\usepackage{amsmath}
\usepackage{amssymb}
\usepackage{mathrsfs}
\usepackage{graphicx}
\usepackage[colorlinks=true, letterpaper=true, pdfstartview=FitV, linkcolor=blue, citecolor=blue, urlcolor=blue]{hyperref}

\begin{document}

\title{Highly sensitive temperature sensing via quadratic optomechanical coupling}
\author{Yu-Sheng Tang}
\affiliation{Key Laboratory of Low-Dimensional Quantum Structures and
Quantum Control of Ministry of Education, Key Laboratory for Matter
Microstructure and Function of Hunan Province, Department of Physics and
Synergetic Innovation Center for Quantum Effects and Applications, Hunan
Normal University, Changsha 410081, China}

\author{Xun-Wei Xu}
\email{xwxu@hunnu.edu.cn}
\affiliation{Key Laboratory of Low-Dimensional Quantum Structures and
Quantum Control of Ministry of Education, Key Laboratory for Matter
Microstructure and Function of Hunan Province, Department of Physics and
Synergetic Innovation Center for Quantum Effects and Applications, Hunan
Normal University, Changsha 410081, China} 
\affiliation{Hunan Research Center of the Basic Discipline for Quantum Effects and Quantum Technologies, Hunan Normal University, Changsha 410081, China}
\affiliation{Institute of Interdisciplinary Studies, Hunan Normal University, Changsha, 410081, China}

\author{Jie-Qiao Liao}
\affiliation{Key Laboratory of Low-Dimensional Quantum Structures and
Quantum Control of Ministry of Education, Key Laboratory for Matter
Microstructure and Function of Hunan Province, Department of Physics and
Synergetic Innovation Center for Quantum Effects and Applications, Hunan
Normal University, Changsha 410081, China} 
\affiliation{Hunan Research Center of the Basic Discipline for Quantum Effects and Quantum Technologies, Hunan Normal University, Changsha 410081, China}
\affiliation{Institute of Interdisciplinary Studies, Hunan Normal University, Changsha, 410081, China}

\author{Hui Jing}
\affiliation{Key Laboratory of Low-Dimensional Quantum Structures and
Quantum Control of Ministry of Education, Key Laboratory for Matter
Microstructure and Function of Hunan Province, Department of Physics and
Synergetic Innovation Center for Quantum Effects and Applications, Hunan
Normal University, Changsha 410081, China} 
\affiliation{Hunan Research Center of the Basic Discipline for Quantum Effects and Quantum Technologies, Hunan Normal University, Changsha 410081, China}
\affiliation{Institute of Interdisciplinary Studies, Hunan Normal University, Changsha, 410081, China}

\author{Le-Man Kuang}
\affiliation{Key Laboratory of Low-Dimensional Quantum Structures and
Quantum Control of Ministry of Education, Key Laboratory for Matter
Microstructure and Function of Hunan Province, Department of Physics and
Synergetic Innovation Center for Quantum Effects and Applications, Hunan
Normal University, Changsha 410081, China} 
\affiliation{Hunan Research Center of the Basic Discipline for Quantum Effects and Quantum Technologies, Hunan Normal University, Changsha 410081, China}
\affiliation{Institute of Interdisciplinary Studies, Hunan Normal University, Changsha, 410081, China}

\begin{abstract}
The effective frequency of a mechanical resonator can be tuned via the spring effect induced by quadratic optomechanical (QOM) coupling,
and both spontaneous symmetry breaking and anti-parity-time phase transition were predicted in the QOM systems.
Here, we show that the mechanical susceptibility can be enhanced significantly by driving the QOM system with a strong external optical field, and divergence will happen as the driving strength approaches the critical point (CP) for spontaneous symmetry breaking.
Based on the CP, we propose a highly sensitive temperature sensor with a mechanical resonator quadratically coupled to an optical mode.
We find that the sensitivity of the temperature sensor can be enhanced by several orders of magnitude as the driving strength approaches the CP, and the sensitivity of the temperature sensor remains high in the low-temperature limit.
Our work provides an effective way to realize highly sensitive temperature sensing at ultra-low temperature in the QOM systems.
\end{abstract}

\maketitle

\section{Introduction}

Optomechanical systems~\cite{Aspelmeyer2014RMP}, based on the parametric coupling between
optical and mechanical resonators, provide a suitable platform for both
macroscopic quantum effects exploration~\cite{Aspelmeyer2012PT,Liu_2013CPB,Liu2018CPB} and quantum technology applications~\cite{Barzanjeh2022NatPh,Metcalfe2014ApPRv,LiBB2021Nanop}.
The optomechanical interaction for the frequency of the cavity depending on the amplitude of
mechanical motion has enabled precision sensing of various physical
quantities, including displacements~\cite{Arcizet2006PRL,Rocheleau_2009Nat,Whittle2021Sci,Magrini2021Natur,Zhou2023NatCo}, masses~\cite{LI2013223PR,YuW2016NatCo,LinQ2017PhRvA,Sansa2020NatCo,WENG202350}, forces~\cite{LiM2008Natur,Teufel2009NatNa, Gavartin2012NatNa,Schreppler2014Sci,Weber2016NatCo,Reinhardt2016PRX,Ahn2020NatNa,Fogliano2021NatCo},
accelerations~\cite{Krause2012NaPho,Guzman2014APL}, magnetic fields~\cite{Rugar2004Natur,Forstner2012PRL,Wu2017NatNa,LiBB2018Optica,Colombano2020PRL}, and ultrasounds~\cite{Basiri2019NatCo,Westerveld2021NaPho}.

If there is no linear optomechanical (LOM) coupling, i.e., the first derivative of the cavity frequency versus the mechanical displacement is zero, then we should consider the nonlinear optomechanical coupling, e.g., quadratic optomechanical (QOM) coupling for the cavity frequency depending on the square of the mechanical displacement. In addition to the interesting phenomena already found by linear optomechanical coupling~\cite{Nunnenkamp2010PRA,Huang2011PRA,Tan2013PRA,Xuereb2013PRA,Shi2013PRA,Asjad2014PRA,Liao2014NatSR,Xie2016PRA,SiLG2017PRA,LuXY2018PRAPP,LiuSP2019PRA,Zhang2020OE},
qualitatively novel quantum effects are expected by QOM coupling, including quantum nondemolition
measurements of phonon number~\cite{ThompsonNat08,Jayich2008NJP,Miao2009PRL,Ludwig2012PRL}, measurement of
phonon shot noise~\cite{Clerk2010PRL,dumont2022arxiv}, phononic Josephson oscillation~\cite{XuXW2017PRA}, quantum nonreciprocality~\cite{Xu2020PRJ}, and zeptonewton force sensing~\cite{ZhangSD2022arXiv}.

QOM systems can create tunable adiabatic double-well potentials for mechanical elements, which render the possibilities for the observation of mechanical macroscopic tunneling~\cite{Buchmann2012PRL}, mechanical spontaneous symmetry breaking and pitchfork bifurcation~\cite{Larson2011PRA,Seok2013PRA,Seok2014PRA,Ruiz2016PRA,Wurl2016PRA,Xu2024NatNa}. First- and second-order buckling transitions between stable mechanical states have been reported experimentally with a dielectric membrane in the middle of the optical cavity~\cite{XuH2017NatCo}.
QOM interaction can also induce exceptional points (EPs)~\cite{Xu2022Arxiv,Djorwe2019PRAPP}, corresponding to the transitions between anti-parity-time broken (Anti-PTB) and anti-parity-time symmetry (Anti-PTS).
Various phase transitions were predicted in QOM systems, and the phase transitions can be used to enhance the sensitivity for sensing in the QOM systems.

Temperature is one of the fundamental physical parameters with important applications in various physical, chemical, and biological processes.
Besides the traditional resistance thermometry~\cite{Giazotto2006RMP}, numerous new technologies and devices for thermal sensing have been developed in recent years~\cite{Dedyulin2022MeScT,Moreva2020PRAPP,2021NatCo,WangJ2020PRR,Brenes2023PRA}, including photonic thermometry~\cite{Xu2014OE,KLIMOV2018308,Liao2021LSA,DongCH2009APL,LiBBAPL2010,XieD2024PRR} and optomechanical thermometry~\cite{Purdy2015PRA,WangQ2015PRA,Purdy2017Sci,Chowdhury2019QS,Montenegro2020PRR,Galinskiy2020Optica,Singh2020PRL,Shirzad2024PRA},
and quantum thermometry is concerned with finding the fundamental limits on the accuracy of temperature measurements in the quantum regime~\cite{Mehboudi2019JPhA,Potts2020PRAPP,Mok2021CmPhy,ZhangN2022PhRvP,Kuang2023PRA2,TanQS2024PRA,Aiache2024PRE,ZhangN2024PRA,Aiache2024PRA,Zhang2022npjQI,Mukherjee2019CmPhy,Ullah2023,Ullah2025}.
In particular, the thermometry becomes exponentially inefficient at low temperatures~\cite{Correa2017PRA,Hovhannisyan2018PRA,Mehboudi2019PRL,Kuang2023PRA1}.
Here, we will show how to enhance the sensitivity for temperature sensing at ultra-low temperature via QOM coupling.
Optomechanical systems have unique advantages in temperature sensing due to the mechanical resonance enhanced thermal sensing and optical resonance enhanced readout sensitivity~\cite{Purdy2015PRA,WangQ2015PRA,Purdy2017Sci,Chowdhury2019QS,Montenegro2020PRR,Galinskiy2020Optica,Singh2020PRL}.

We find that the mechanical susceptibility can be enhanced significantly by driving the QOM system with a strong external optical field, and divergence will happen as the driving strength approaches the critical point (CP) for spontaneous symmetry breaking, which provides an excellent mechanism for highly-sensitive measurement.
As an example, we show that the QOM system with a strong external optical field can be used to achieve temperature sensing. Interestingly, the sensitivity for temperature sensing can be enhanced by several orders with the driving strength near the CP, and the performance of the temperature sensor does not decline in the low-temperature limit, which can be applied to achieve highly-sensitive temperature sensing at ultra-low temperature.

The remainder of this paper is organized as follows. In Sec.~\ref{QOM_Model}, we introduce the theoretical model of a QOM system, and derive the power spectral density (PSD) of the mechanical resonator. The mechanical susceptibility for varying driving strength is studied in Sec.~\ref{MS}. In Sec.~\ref{TS}, we show the performance of a temperature sensor based on QOM coupling. Finally, the conclusions and some discussions are given in Sec.~\ref{DC}.

\section{Quadratic optomechanics}\label{QOM_Model}

\begin{figure}[tbp]
\includegraphics[bb=47 227 567 654, width=8.5 cm, clip]{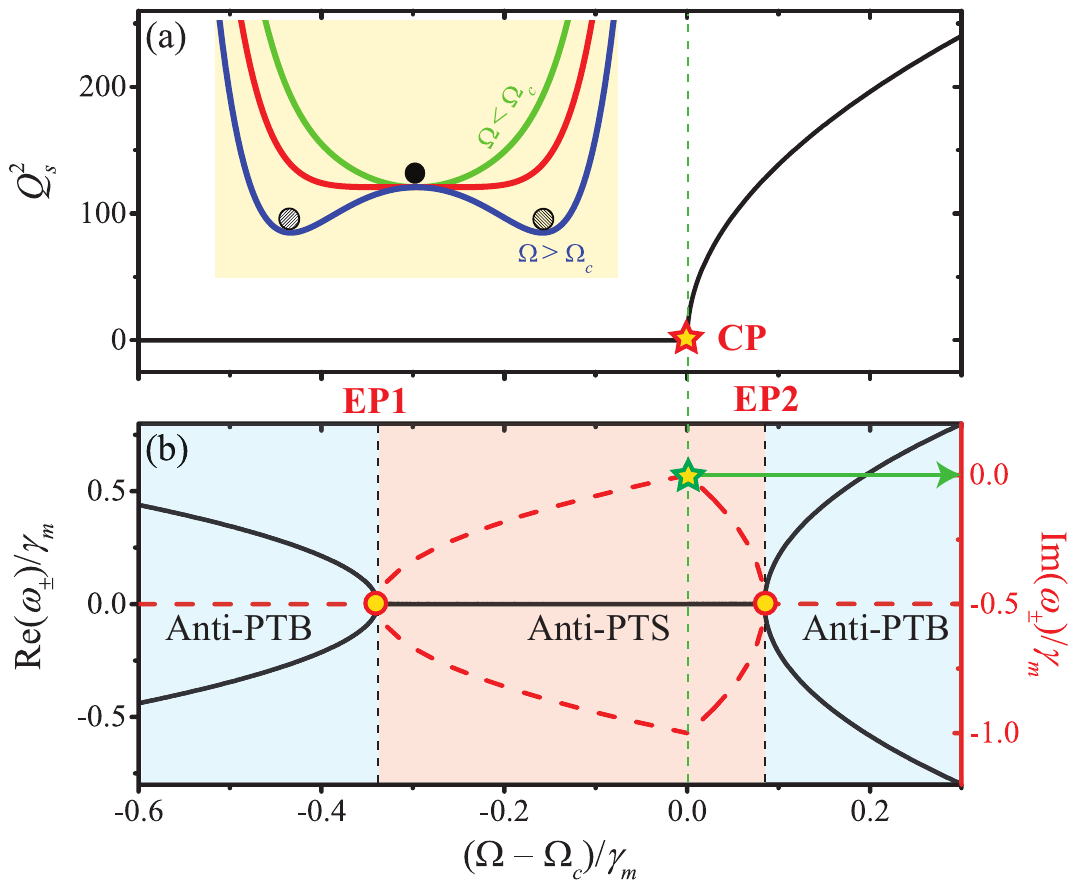}
\caption{(Color online) (a) The steady-state value of the square of the mechanical position $Q_{s}^{2}$ is plotted as a function of the coupling strength $\Omega$ [inset, the effective potential functions for $\Omega<\Omega_c$ (green), $\Omega=\Omega_c$ (red), and $\Omega>\Omega_c$ (blue)]. (b) The real (solid, left axis) and imaginary (dashed, right axis) parts of $\omega_{\pm}$ are shown as functions of $\Omega$. The other parameters are $\protect\gamma_c/2\protect\pi= 5$ GHz, $\protect%
\omega_m/2\protect\pi= 8.7$ MHz, $\omega_m/\gamma_m= 10^4$, and $g/2\protect\pi=- 245$ Hz.}
\label{fig1}
\end{figure}

We consider a QOM system that an optical mode couples to a mechanical mode in a quadratic fashion.
In experiments, the QOM coupling has been explored in various cavity-optomechanical systems, including
membrane in the middle of Fabry-Perot cavities~\cite{ThompsonNat08,Sankey2010NatPh,Karuza2012JOP,LeeD2015NC},
nanosphere levitated in Fabry-Perot cavities~\cite{Fonseca2016PRL,Uros2019PRL,Bullier2021PRR},
cold atoms trapped in Fabry-Perot cavities~\cite{Purdy2010PRL},
double microdisk resonators~\cite{Hill2013PHD},
microdisk-nanocantilever systems~\cite{Doolin2014PRA},
microsphere-nanostring systems~\cite{Brawley2016NatCo},
paddle nanocavities~\cite{Kaviani2015Optica}
and double-slotted photonic crystal cavities~\cite{ParaisoPRX15,Burgwal2023NatCo}.
QOM model also can be simulated in a superconducting electrical circuit~\cite{Kim2015PRA,Eichler2018PRA,Zhou2021OE,Yin2022PRA}, which leads to the possibility of achieving quadratic optomechanics in the single-photon strong-coupling regime.
In addition, a purely QOM system (i.e., no LOM coupling) can be realized by displacing the mechanical resonator at a quadratic point~\cite{dumont2022arxiv}, so that no net radiation force is applied on it (otherwise there
would necessarily be LOM coupling).

In this paper, we consider a purely QOM system for a single optical mode quadratic coupling to a mechanical resonator, by adiabatically eliminating the other optical modes for the separation of eigenfrequencies~\cite{dumont2022arxiv}.
The optical mode is driven resonantly by an external field at a frequency $ \omega _{L}$.
In the rotating reference frame with frequency $\omega _{L}$, the Hamiltonian of the QOM system reads
\begin{equation}
\frac{H}{\hbar}=\frac{1}{2}\omega _{m}\left( Q^{2}+P^{2}\right) +2gA^{\dag }AQ^{2}+\Omega
\left( A^{\dag }+A\right).
\end{equation}%
Here $A$ and $A^{\dag }$ are the annihilation and creation operators of the
optical mode with resonant frequency $\omega _{c}=\omega _{L}$; $Q$ and $P$ are the
dimensionless displacement and momentum operators of the mechanical mode
with frequency $\omega _{m}$; $g$ is the single-photon quadratic optomechanical
coupling coefficient.
We assume that the coupling strength of the optical
driving field $\Omega$ is much stronger than the damping rates of the optical and mechanical modes ($
\gamma _{c}$ and $\gamma _{m}$). i.e., $\Omega \gg \{\gamma _{c},\:\gamma _{m}\}$.

The dynamics of the QOM system can be determined by the quantum Langevin equations~\cite{Gardiner2000QN}
\begin{equation} \label{QLE1}
\frac{dQ}{dt}=\omega _{m}P,
\end{equation}%
\begin{equation} \label{QLE2}
\frac{dP}{dt}=-\omega _{m}Q-4gA^{\dag }AQ-\gamma _{m}P+\xi ,
\end{equation}
\begin{equation} \label{QLE3}
\frac{dA}{dt}=-\frac{\gamma _{c}}{2}A-i2gAQ^{2}-i\Omega +\sqrt{\gamma _{c}}%
a_{\mathrm{in}}.
\end{equation}%
Here, $a_{\mathrm{in}}(t)$ is the optical noise operator satisfying the correlation function
\begin{equation}
\left\langle a_{\mathrm{in}}\left( t\right) a_{\mathrm{in}}^{\dag }\left(
t^{\prime }\right) \right\rangle =\delta \left( t-t^{\prime }\right),
\end{equation}%
and $\xi(t)$ denotes the quantum Brownian force satisfying the correlation function
\begin{equation}
\left\langle \xi \left( t\right) \xi \left( t^{\prime }\right) \right\rangle
=\frac{\gamma _{m}}{\omega _{m}}\int \frac{d\omega }{2\pi }\omega
e^{-i\omega \left( t-t^{\prime }\right) }\left[ 1+\coth \left( \frac{\hbar
\omega }{2k_{B}T}\right) \right],
\end{equation}%
where $k_B$ is the Boltzmann constant and $T$ is the temperature of the environment.

Under the mean-field approximation, the steady-state solution of the quantum Langevin equations~(\ref{QLE1})-(\ref{QLE3}) can be obtained by
first replacing the operators by the mean values ($\langle A\rangle =\alpha$, $\langle Q\rangle =Q_s$ and $\langle P\rangle = P_{s}$), and then setting the time derivatives to zeros.
If the quadratic optomechanical coupling is positive, i.e., $g>0$, we have the steady-state values
\begin{equation}
\alpha =-i\frac{2\Omega }{ \gamma _{c} },
\end{equation}%
and 
\begin{equation}
P_{s} =Q_{s}=0.
\end{equation}%
For a negative quadratic optomechanical coupling, i.e., $g<0$, the steady-state values are obtained as
\begin{equation}
\alpha =\frac{-i2\Omega }{\left( \gamma _{c}+i4gQ_{s}^{2}\right) },
\end{equation}%
\begin{equation}
P_{s} =0,
\end{equation}%
\begin{equation}
Q_{s}^{2}=\left\{
\begin{array}{cc}
0 & \Omega \leq \Omega _{c} \\
\sqrt{-\frac{1}{4g^{2}}\left[ \frac{4g\Omega ^{2}}{\omega _{m}}+\left( \frac{%
\gamma _{c}}{2}\right) ^{2}\right] } & \Omega >\Omega _{c}%
\end{array}%
\right. ,
\end{equation}
with a critical driving strength
\begin{equation}
\Omega _{c}\equiv \sqrt{-\frac{\gamma _{c}^{2}\omega _{m}}{16g}}.
\end{equation}
In the following, we will focus on the case of $g<0$.
The steady-state value of the square of the mechanical position $Q_{s}^{2}$ is plotted as a function of the coupling strength $\Omega$ in Fig.~\ref{fig1}(a).
The figure predicts a spontaneous symmetry breaking at the critical strength $\Omega_c$, i.e., the CP.
The inset shows the effective potential functions for $\Omega<\Omega_c$ (green), $\Omega=\Omega_c$ (red), and $\Omega>\Omega_c$ (blue).
Similar phenomenon has been discussed in detail in Refs.~\cite{Seok2013PRA,Seok2014PRA,Ruiz2016PRA}.

We are interested in the fluctuation spectra induced by the noises. All the operators can be rewritten as the sum of their steady-state values and quantum
fluctuations: $A\rightarrow \alpha +a$, $Q\rightarrow Q_{s}+q$, and $%
P\rightarrow P_{s}+p$, then quantum Langevin equations for the quantum flucturation operators $a$, $q$, and $p$ are given by
\begin{equation}
\frac{dq}{dt}=\omega _{m}p,
\end{equation}%
\begin{equation}
\frac{dp}{dt}=-\left( \omega _{m}+4g\left\vert \alpha \right\vert
^{2}\right) q-4gQ_{s}\left( \alpha ^{\ast }a+\alpha a^{\dag }\right) -\gamma _{m} p+\xi,
\end{equation}%
\begin{equation}\label{LE3}
\frac{da}{dt}=\left( -\frac{\gamma _{c}}{2}-i2gQ_{s}^{2}\right) a-i4g\alpha
Q_{s}q+\sqrt{\gamma _{c}}a_{\mathrm{in}}.
\end{equation}%
By introducing the Fourier transform%
\begin{equation}
f\left( \omega \right) =\frac{1}{\sqrt{2\pi }}\int_{-\infty }^{+\infty
}f\left( t\right) e^{i\omega t}dt,
\end{equation}%
the quantum Langevin equations can be solved analytically in the frequency domain.
Specifically, the position flucturation operator is obtained as
\begin{equation}
q\left( \omega \right) =\chi \left( \omega \right) \left\{\xi \left( \omega \right)
+\eta \left( \omega \right) a_{\mathrm{in}}\left(
\omega \right) + \left[ \eta \left( -\omega
\right) \right] ^{\ast }a_{\mathrm{in}}^{\dag }\left( \omega \right)\right\}
\end{equation}%
with mechanical susceptibility
\begin{equation}\label{MS1}
\chi \left( \omega \right) \equiv \frac{\omega _{m}}{\omega _{m}^{2}-\omega
^{2}-i\gamma _{m}\omega +4\omega _{m}g\left\vert \alpha
\right\vert ^{2}\zeta \left( \omega \right) },
\end{equation}%
and
\begin{equation}
\eta \left( \omega \right) \equiv \frac{-4gQ_{s}\alpha ^{\ast }\sqrt{\gamma
_{c}}}{\left[ \frac{\gamma _{c}}{2}+i\left( 2gQ_{s}^{2}-\omega
\right) \right] },
\end{equation}%
\begin{equation}
\zeta \left( \omega \right) \equiv 1-\frac{\left( 4gQ_{s}^{2}\right) ^{2}}{%
\left( \frac{\gamma _{c}}{2}-i\omega \right) ^{2}+\left( 2gQ_{s}^{2}\right)
^{2}}.
\end{equation}

The temperature of the environment can be measured by the PSD of the mechanical resonator defined by $S_{qq}\left( \omega \right) \equiv \int d\omega ^{\prime
}\left\langle q\left( \omega \right) q\left( \omega ^{\prime }\right)
\right\rangle $, which reads
\begin{equation}
S_{qq}\left( \omega \right) =\left\vert \chi \left( \omega \right)
\right\vert ^{2}\left[ S_{m,\mathrm{th}}\left( \omega \right) +S_{c,\mathrm{%
vac}}\left( \omega \right) \right],
\end{equation}%
where%
\begin{equation}
S_{m,\mathrm{th}}\left( \omega \right) =\frac{\gamma _{m}}{\omega
_{m}}\omega \left[ 1+\coth \left( \frac{\hbar \omega }{2k_{B}T}\right) \right]
\end{equation}
is the mechanical thermal noise spectrum, and
\begin{equation}
S_{c,\mathrm{vac}}\left( \omega \right) =\frac{16g^{2}Q_{s}^{2}\left\vert
\alpha \right\vert ^{2}\gamma _{c}}{\left( \frac{\gamma _{c}}{2}\right)
^{2}+\left(2gQ_{s}^{2}-\omega \right) ^{2}}
\end{equation}
is the radiation-induced noise spectrum.
The radiation-induced noise spectrum $S_{c,\mathrm{vac}}\left( \omega \right)$ is irrelevant with the temperature of the environment. Because the frequency of the optical mode is assumed to be high enough so that the thermal photon effect can be neglected.
The signal of the environment is enclosed in mechanical thermal noise spectrum $S_{m,\mathrm{th}}$, but $S_{m,\mathrm{th}}$ does not depend on the QOM coupling.
In other words, the sensitivity to the temperature can only be tuned by adjusting the mechanical susceptibility $\chi \left( \omega \right)$ via QOM coupling.

Let us do some discussion on the mechanical susceptibility in the next section first.
Without loss of generality, in the following calculations, we will take the experimental parameters~\cite{ParaisoPRX15}: QOM coupling strength $g/2\pi=-245$ Hz, mechanical resonance frequency $\omega _{m}/2\pi=8.7$ MHz, mechanical quality factor $Q_{m} = 10^4$, and optical damping rates $\gamma _{c}/2\pi = 5$ GHz, i.e., the system works in the sideband unresolved regime with $\gamma _{c}\gg \omega _{m}$.

\section{Mechanical susceptibility}\label{MS}

\begin{figure}[tbp]
\includegraphics[bb=48 132 513 629, width=8.5 cm, clip]{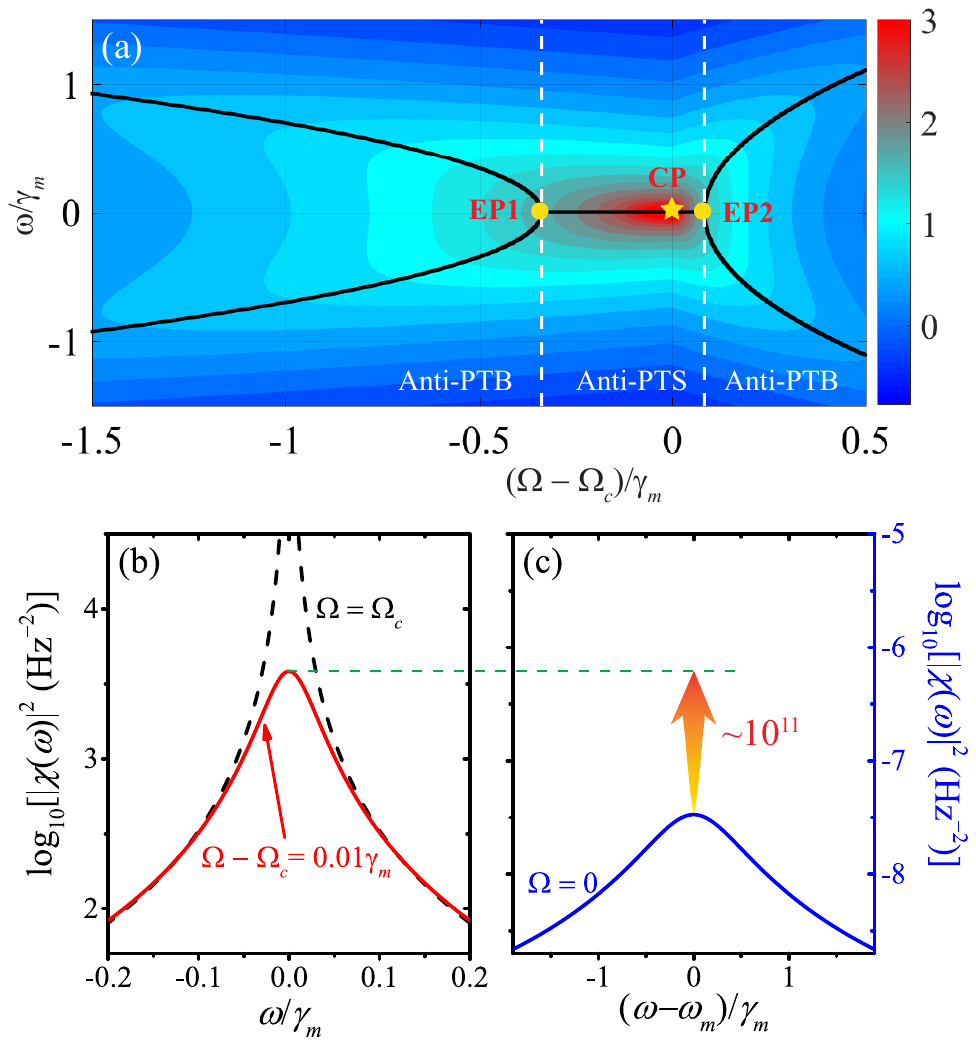}
\caption{(Color online) (a) Mechanical susceptibility $%
\log_{10}[|\chi (\omega)|^2]$ for varying frequency $\protect\omega$
and driving strength $\Omega-\Omega_c$. The black solid curves represent the
real parts of $\omega _{\pm }$ given in Eq.~(\ref{eigen1}). $\log_{10}[|\chi (\omega)|^2]$ is
plotted as a function of frequency $\protect\omega$ in (b) and (c): (b) black dashed curve for $%
\Omega = \Omega_c$ and red solid curve for $%
\Omega - \Omega_c = 0.01 \gamma_m$; (c) blue solid curve for $\Omega=0$. The other parameters are $\protect\gamma_c/2\protect\pi= 5$ GHz, $\protect%
\omega_m/2\protect\pi= 8.7$ MHz, $\omega_m/\gamma_m= 10^4$, and $g/2\protect\pi=- 245$ Hz.}
\label{fig2}
\end{figure}

Mechanical susceptibility plays a critical role in the temperature sensing based on the PSD of a mechanical resonator~\cite{XuA2022PRA}.
The mechanical susceptibility in Eq.~(\ref{MS1}) is adjusted by the QOM coupling via the last term in the denominator, i.e., $4\omega _{m}g\left\vert \alpha
\right\vert ^{2}\zeta \left( \omega \right)$, where $\zeta \left( \omega \right)$ is frequency dependent.
However, in the certain regime $\omega <\omega _{m}\ll \frac{\gamma _{c}}{2}$ we will focus on, it can be given approximatively as
\begin{equation}
\zeta \left( \omega \right) \approx \zeta _{0}\equiv 1-\frac{\left(
4gQ_{s}^{2}\right) ^{2}}{\left( \frac{\gamma _{c}}{2}\right) ^{2}+\left(
2gQ_{s}^{2}\right) ^{2}}.
\end{equation}%
Then mechanical susceptibility can be rewritten as
\begin{equation}
\chi \left( \omega \right)  =\frac{-\omega _{m}}{\left( \omega -\omega
_{+}\right) \left( \omega -\omega _{-}\right) }
\end{equation}%
with
\begin{equation}\label{eigen1}
\omega _{\pm }=-i\frac{\gamma _{m}}{2}\pm \sqrt{\omega _{m}\left( \omega
_{m}+4g\left\vert \alpha \right\vert ^{2}\zeta _{0}\right) -\left( \frac{%
\gamma _{m}}{2}\right) ^{2}}.
\end{equation}%
Clearly, the properties of mechanical susceptibility $\chi \left( \omega \right)$ are determined by the values of $\omega _{\pm }$.

The real parts of $\omega_{\pm}$ are shown as functions of $\Omega$ in Fig.~\ref{fig1}(b).
Interestingly, there are two exceptional points (EP1 and EP2) for $\omega_{+}=\omega_{-}$, with the coupling strength
\begin{equation}
\Omega _{\mathrm{EP1}}^{2}=\Omega _{c}^2\left[1-\left(\frac{\gamma_m}{2\omega_m}\right)^2\right],
\end{equation}%
and
\begin{equation}
\Omega _{\mathrm{EP2}}^{2}=\Omega _{c}^2\left[1-\left(\frac{\gamma_m}{4\omega_m}\right)^2\right]^{-1},
\end{equation}%
that divide the parameters into three regions: Anti-PTB, Anti-PTS, and Anti-PTB (see Ref.~\cite{Xu2022Arxiv} for more details).
In the Anti-PTS regime, there is a CP for one of the imaginary parts of $\omega_{\pm}$ equals to zero, which corresponds to the spontaneous symmetry breaking at the critical coupling strength $\Omega_c$ as shown in Fig.~\ref{fig1}(a).

The mechanical susceptibility $\log_{10} [|\chi(\protect\omega)|^2]$ for varying frequency $\protect\omega$
and driving strength $\Omega-\Omega_c$ is shown in Fig.~\ref{fig2}(a).
The frequencies of the peaks in the mechanical susceptibility are entirely consistent with the real parts of $\omega _{\pm }$ (black solid curves) given in Eq.~(\ref{eigen1}).
When the driving strength $\Omega$ is in the Anti-PTB regimes, i.e., $\Omega<\Omega _{\mathrm{EP1}}$ or $\Omega>\Omega _{\mathrm{EP2}}$, there are two peaks in the $|\chi(\protect\omega)|^2$; while in the Anti-PTS regime, there is only one peak around the frequency $\omega=0$ and the height of the peak is much higher than that of the two peaks.
The height of the peak becomes even higher as the driving strength $\Omega$ coming closer to the critical strength $\Omega_c$, and divergence will happen as the driving strength $\Omega$ approaches the critical strength $\Omega_c$ [dashed curve in Fig.~\ref{fig2}(b)].
The divergence of $\chi(\protect\omega)$ is induced by the zero dissipation (i.e., ${\rm Im}(\omega_{\pm})=0$) at the CP as $\Omega=\Omega_c$, so that zero appears in the denominator of the mechanical susceptibility.
For a comparison, the mechanical susceptibility $\log_{10} [|\chi(\protect\omega)|^2]$ without optical external driving $\Omega=0$, i.e., the QOM effect can be ignored, is shown in Fig.~\ref{fig2}(c).
So surprisingly, the mechanical susceptibility $|\chi(\protect\omega)|^2$ can be enhanced by eleven orders of magnitude for $\Omega= \Omega_c + 0.01 \gamma_m$ as compared with the case for $\Omega=0$.
As the PSD of a mechanical resonator is linearly correlated with the mechanical susceptibility, the enhancement of the mechanical susceptibility can enable opportunities for high sensitivity sensing based on detection of PSD, such as temperature sensing.

To avoid low-frequency electronic noise in the measurements, the mechanical oscillator’s PSD can be extracted by the heterodyne
quadrature measurement.
In addition, the PSD of the mechanical resonator can also be extracted by measuring the power spectrum of the optical mode in two different ways.
One way is applying another probe optical laser. A weak probe optical beam bypass the cavity with low reflectivity from the cavity mirror but high reflectivity from the mechanical resonator, and the reflection of probe field from the mechanical resonator can be sent to a spectrum analyzer for the power spectrum measurements~\cite{JSheng2020PRL}.
The other way is measuring the power spectrum of the optical mode directly. 
Specifically, from Eq.~(\ref{LE3}), the amplitude and phase quadratures ($X\equiv (a+a^{\dag})/\sqrt{2}$, $Y\equiv i(a^{\dag}-a)/\sqrt{2}$) of the optical mode in the frequency domain are given by
\begin{eqnarray}\label{EQ29}
X\left( \omega \right)  &=&\left\{ \left[ \mu \left( -\omega \right) \right]
^{\ast }+\mu \left( \omega \right) \right\} q\left( \omega \right)  
\nonumber \\
&&+\upsilon \left( \omega \right) a_{\mathrm{in}}^{\dag }\left( \omega
\right) +\left[ \upsilon \left( -\omega \right) \right] ^{\ast }a_{\mathrm{in%
}}\left( \omega \right) ,
\end{eqnarray}%
\begin{eqnarray}\label{EQ30}
Y\left( \omega \right)  &=&\left\{ i\mu \left( \omega \right) -i\left[ \mu
\left( -\omega \right) \right] ^{\ast }\right\} q\left( \omega \right)  
\nonumber \\
&&-i\left[ \upsilon \left( -\omega \right) \right] ^{\ast }a_{\mathrm{in}%
}\left( \omega \right) +i\upsilon \left( \omega \right) a_{\mathrm{in}%
}^{\dag }\left( \omega \right) ,
\end{eqnarray}
with
\begin{equation}
\mu \left( \omega \right) =\frac{i2\sqrt{2}g\alpha ^{\ast }Q_{s}}{\left[
\left( \gamma _{c}/2\right) -i\left( 2gQ_{s}^{2}+\omega \right) \right] },
\end{equation}%
\begin{equation}
\upsilon \left( \omega \right) =\frac{\sqrt{\gamma _{c}/2}}{\left[ \left(
\gamma _{c}/2\right) -i\left( 2gQ_{s}^{2}+\omega \right) \right] }.
\end{equation}%
From Eqs.~(\ref{EQ29}) and (\ref{EQ30}), we find that the amplitude and phase quadratures ($X\left( \omega \right)$ and $Y\left( \omega \right)$) are related to the mechanical position $q\left( \omega \right)$. If $\mu \left( \omega \right)\neq 0$, the PSD of the mechanical resonator $S_{qq}\left( \omega \right)$ can be obtained by measuring the PSD of the optical mode.
In the regime of $\left( 2gQ_{s}^{2}\pm\omega \right) \ll \gamma _{c}$, we have $\mu \left( \omega \right) \approx i4\sqrt{2}g\alpha ^{\ast }Q_{s}/\gamma _{c}$.
The condition of $\mu \left( \omega \right)\neq 0$ is satisfied in the regime of $\Omega > \Omega_c$ for $Q_{s}\neq 0$.

\section{Temperature sensing} \label{TS}

\begin{figure}[tbp]
\includegraphics[bb=170 411 401 646, width=8.5 cm, clip]{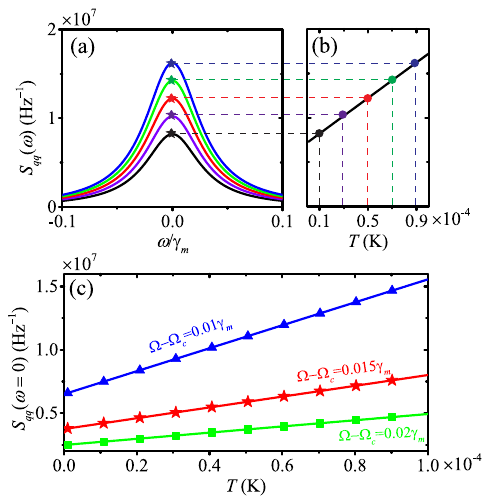}
\caption{(Color online) Working principle of optomechanical temperature
sensing via QOM coupling. (a) Mechanical position power spectral densities $S_{qq}(%
\protect\omega)$ as a function of frequency $\protect\omega$. (b) The peak
values of $S_{qq}(\protect\omega =0)$ versus the temperature $T$. (c) $%
S_{qq}(\protect\omega =0)$ versus the temperature $T$ for different driving
strengths $(\Omega-\Omega_c)/\protect\gamma_m=0.01,\,0.015,\,0.02$. The other parameters are the
same as in Fig.~\protect\ref{fig2}.}
\label{fig3}
\end{figure}

\begin{figure}[tbp]
\includegraphics[bb=120 265 458 614, width=8.5 cm, clip]{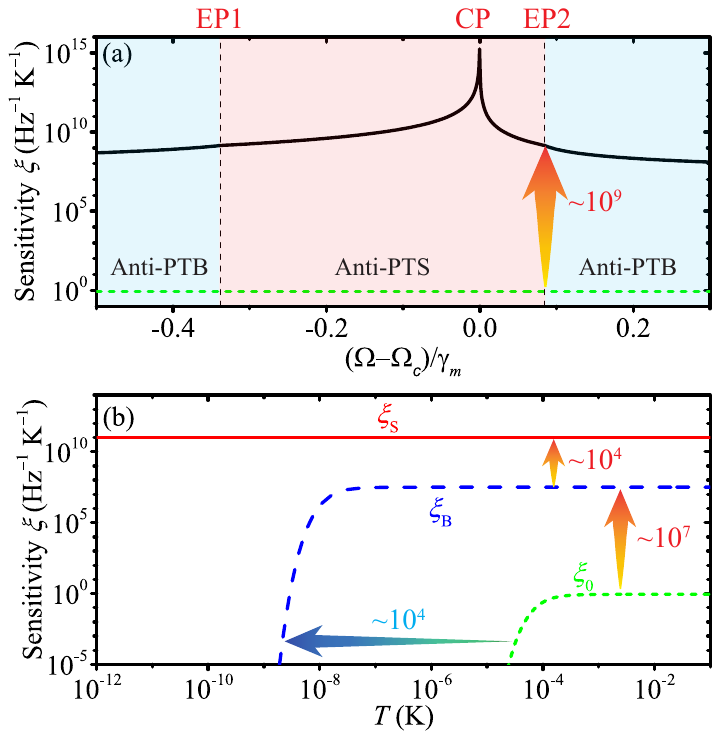}
\caption{(Color online) (a) Sensitivity $\xi$ (solid red curve) versus driving strengths $%
(\Omega-\Omega_c)/\protect\gamma_m$ in high temperature limit.
Sensitivity $\xi_0$ for the case of no driving $\Omega=0$ is shown by green short-dashed curve.
(b) Sensitivity $\xi$ versus the temperature $T$ in different regimes: red solid curve for $\xi_{\rm S}$ with $%
(\Omega-\Omega_c)= 0.01 \gamma_m$ in Anti-PTS regime, blue dashed curve for $\xi_{\rm B}$ with $%
(\Omega-\Omega_c)= \gamma_m$ in Anti-PTB regime, green short-dashed curve for $\xi_0$ without driving $\Omega=0$. The other parameters are the same as in Fig.~\protect\ref{fig2}.}
\label{fig4}
\end{figure}

In this section, we will reveal the relationship between the PSD and the temperature of the environment and discuss the sensitivity for the QOM system working as a temperature sensor.
As expected, the height of the peak in the PSD depends on temperature of the mechanical resonators, as shown in Fig.~\ref{fig3}(a).
More specifically, in the Anti-PTS regime, there is only one peak at $\omega=0$ and the height of the peak increases linearly with temperature, as shown in Fig.~\ref{fig3}(b).
The derivative of the height with respect to the temperature, i.e., the slope of the curve, depends on the driving strength $\Omega$ of the external fields.
The slope increases quickly as the driving strength $\Omega$ approaches to the critical strength $\Omega_c$, see Fig.~\ref{fig3}(c).

The sensitivity of the QOM temperature sensor can be described by the slope of the PSD as
\begin{equation}
\xi(\omega) \equiv \frac{d S_{qq}(\omega)}{dT}.
\end{equation}
In the Anti-PTS regime at frequency $\omega= {\rm Re}(\omega_{\pm})=0$ , it can be expressed analytically as
\begin{equation}
\xi _{\rm S}=\frac{2k_{B}\omega _{m}\gamma _{m}}{\hbar \left\vert
\omega _{+}\omega _{-}\right\vert ^{2}},
\end{equation}
where $\omega _{\pm}$ given in Eq.~(\ref{eigen1}) are pure imaginary numbers in the Anti-PTS regime.
Clearly, $\xi_{\rm S}$ is temperature independent, and divergence will happen for $\omega _{+}=0$ or $\omega _{-}=0$ (i.e., no dissipation) when the driving strength $\Omega$ approaches the critical strength $\Omega_c$, as shown in Fig.~\ref{fig4}(a).

The sensitivity of the QOM temperature sensor in the Anti-PTB regime can be defined by the slope of the PSD at frequency $\omega_{\rm eff}= {\rm Re}(\omega_{\pm})$ as,
\begin{equation}
\xi _{\mathrm{B}}=\left\vert \chi \left( \omega _{\mathrm{eff}}\right)
\right\vert ^{2}\frac{\hbar \omega _{\mathrm{eff}}^{2}\gamma _{m}}{2\omega
_{m}k_{B}T^{2}}\left[ \sinh \left( \frac{\hbar \omega _{\mathrm{eff}}}{%
2k_{B}T}\right) \right] ^{-2}.
\end{equation}%
Different from $\xi _{\mathrm{S}}$, $\xi _{\mathrm{B}}$ is temperature dependent.
In the high temperature limit $k_{B}T \gg \hbar \omega _{m}$, we have
\begin{equation}
\xi _{\mathrm{B}}=\left\vert \chi \left( \omega _{\mathrm{eff}}\right)
\right\vert ^{2}\frac{2k_{B}\gamma _{m}}{\hbar \omega _{m}},
\end{equation}%
which is temperature independent.
For the case of no external driving, i.e., $\Omega=0$, we have the sensitivity of the temperature sensing $\xi _{0}$ as,
\begin{equation}
\xi _{\mathrm{0}}=\frac{\hbar \omega _{m}}{2\gamma _{m}k_{B}T^{2}}\left[
\sinh \left( \frac{\hbar \omega _{m}}{2k_{B}T}\right) \right] ^{-2},
\end{equation}%
which can be simplified as
\begin{equation}
\xi _{\mathrm{0}}=\frac{2k_{B}}{\hbar \omega _{m}\gamma _{m}}
\end{equation}
in the high temperature limit.
For comparison, we also show the sensitivity $\xi _{0}$ in Fig.~\ref{fig4}(a).
The sensitivity can be extremely enhanced by the QOM coupling under strong external deriving. For example, the sensitivity at EP2 in Fig.~\ref{fig4}(a) is about $1.4\times 10^9$/(Hz K), a factor of $10^9$ above the sensitivity $\xi _{0}\approx 0.876$/(Hz K) without external driving $\Omega=0$, and the enhancement becomes even higher in the Anti-PTS regime.

QOM coupling also can enhance the performance of the temperature sensor in the low temperature limit $k_{B}T \ll \hbar \omega _{\mathrm{eff}}$.
In the low temperature limit $k_{B}T \ll \hbar \omega _{\mathrm{eff}}$, we have
\begin{equation}
\xi _{\mathrm{B}}=\left\vert \chi \left( \omega _{\mathrm{eff}}\right)
\right\vert ^{2}\frac{2\hbar \omega _{\mathrm{eff}}^{2}\gamma _{m}}{\omega
_{m}k_{B}T^{2}}\exp \left( -\frac{\hbar \omega _{\mathrm{eff}}}{k_{B}T}%
\right),
\end{equation}%
for the external driving in the Anti-PTB regime, and
\begin{equation}
\xi _{\mathrm{0}}=\frac{2\hbar \omega _{m}}{\gamma _{m}k_{B}T^{2}}\exp
\left( -\frac{\hbar \omega _{m}}{k_{B}T}\right)
\end{equation}%
for the case of no external driving, i.e., $\Omega=0$.
Both sensitivities $\xi _{\mathrm{B}}$ and $\xi _{\mathrm{0}}$ decline exponentially towards zero when the temperature tends to zero.
To display the behaviors of the sensitivity in the low temperature limit, the sensitivities versus the temperature $T$ in different regimes are shown in Fig.~\ref{fig4}(b). In the Anti-PTB regime, there is a temperature limit that the sensitivity drops dramatically, but the temperature limit becomes much lower with strong external driving. And most remarkably, in the Anti-PTS regime, QOM temperature sensor can operate with consistently high sensitivity $\xi _{\rm S}$ at any low temperature in principle, because $\omega_{\rm eff}= {\rm Re}(\omega_{\pm})=0$ and $k_{B}T \gg \hbar \omega _{\rm eff}$ are always satisfied in the Anti-PTS regime.
\\

\section{Conclusions}\label{DC}

In summary, we demonstrated that the mechanical susceptibility in the QOM system can be enhanced significantly by a strong external optical driving field, and proposed a highly-sensitive temperature sensor based on this effect.
We found that the sensitivity of the temperature sensor can be enhanced by several orders of magnitude as the external driving strength comes close to the CP for spontaneous symmetry breaking, and the high-sensitivity of the temperature sensor remain unchanged in the low-temperature limit.
The application of the enhanced mechanical susceptibility is not limited to temperature sensing, and may also be applied to achieve highly-sensitive sensing of other physical quantities, such as force and acceleration.

\vskip 2pc \leftline{\bf Acknowledgement}

We thank Huilai Zhang and Jie Wang for fruitful discussions.
This work is supported by the Innovation Program for Quantum Science and Technology (Grant No.~2024ZD0301000).
X.-W.X. is supported by the National Natural Science Foundation of China (NSFC) (Grants No.~12064010, No.~12247105, and No.~12421005),
the Sci-Tech Innovation Program of Hunan Province (Grant No.~2022RC1203), 
and Hunan Provincial Major Sci-Tech Program (Grant No.~2023ZJ1010).
J.-Q.L. was supported in part by NSFC (Grants No. 12175061,
No. 12247105, No. 11935006, and No. 12421005), National
Key R\&D Program of China (Grant No. 2024YFE0102400), and Hunan Provincial Major Sci-Tech Program (Grant No. 2023ZJ1010).
H.J. is supported by the NSFC (Grants No.~11935006 and No.~12421005), the Sci-Tech Innovation Program of Hunan Province (2020RC4047), the National Key R\&D Program (2024YFE0102400), and the Hunan Provincial Major Sci-Tech
Program (2023ZJ1010).
L.-M.K. was supported by NSFC (Grants No.~12247105, No.~12175060, No.~12421005, and No.~11935006), the Sci-Tech Innovation Program of Hunan Province (2020RC4047), and the Hunan Provincial Major Sci-Tech
Program (2023ZJ1010).

\bibliography{ref}

\end{document}